\documentclass[referee]{raa}            

\usepackage{graphicx,times}             
\usepackage{natbib}
\usepackage{amssymb,amsmath}
\usepackage{graphicx}

\bibpunct{(}{)}{;}{a}{}{,}

\usepackage[a4paper=true,pagebackref=true]{hyperref}
\hypersetup{pdftitle = The title of my PDF, pdfauthor = My name, pdfsubject= The subject, pdfkeywords = keyword1 keyword2 keyword3} 
\hypersetup{colorlinks = true, linkcolor = green, anchorcolor = red, citecolor = blue, filecolor = red, pagecolor = red, urlcolor = red}
\begin{document}
   \title{Investigating Sulfur Chemistry in the HD\,163296 disk}

   \volnopage{Vol.0 (20xx) No.0, 000--000}      
   \setcounter{page}{1}          

   \author{Rong Ma
      \inst{1,2}
   \and Donghui Quan
      \inst{3,1} 
   \and Yan Zhou
      \inst{4}
    \and Jarken Esimbek
      \inst{1,5,6}
    \and Dalei Li
      \inst{1,2,6}
    \and Xiaohu Li 
      \inst{1,5,6} 
     \and Xia Zhang
      \inst{1, 6} 
    \and Juan Tuo
      \inst{1,2}
    \and Yanan Feng
      \inst{1,2}
   }

   \institute{Xinjiang Astronomical Observatory, Chinese Academy of Sciences, 150 Science 1-Street, Urumqi, Xinjiang 830011, China; {\it donghui.quan@zhejianglab.com}\\
        \and
            University of Chinese Academy of Sciences, Beijing 100049, China\\
        \and
            Research Center for Intelligent Computing Platforms, Zhejiang Laboratory, Hangzhou 311100, China\\
        \and
           BinZhou University, Huanghe Road, Binzhou City, Shandong, 256600, China\\
        \and
            Key Laboratory of Radio Astronomy, Chinese Academy of Sciences, Urumqi, Xinjiang 830011, People's Republic of China\\
         \and
            Xinjiang Key Laboratory of Radio Astrophysics, 150 Science 1-Street, Urumqi 830011, China
}        
\vs\no
   {\small Received~~20xx month day; accepted~~20xx~~month day}

\abstract{Sulfur chemistry in the formation process of low-mass stars and planets remains poorly understood. The protoplanetary disks (PPDs) are the birthplace of planets and its distinctive environment provides an intriguing platform for investigating models of sulfur chemistry. We analyzed the ALMA observations of CS 7-6 transitions in the HD\,163296 disk and perform astrochemical modeling to explore its sulfur chemistry. We simulated the distribution of sulfur-containing molecules and compared it with \textcolor{black}{observationally deduced fractional
column densities}. We have found that the simulated column density of CS is consistent with the \textcolor{black}{observationally deduced fractional column densities}, while the simulated column density of C$_2$S is lower than the \textcolor{black}{observationally deduced upper limits on column densities}. This results indicate that we have a good understanding of the chemical properties of CS and C$_2$S in the disk. We also investigated the influence of the C/O ratio on sulfur-containing molecules and found that the column densities of SO, SO$_2$, and H$_2$S near the central star are dependent on the C/O ratio. Additionally, we found that the $N$[CS]/$N$[SO] ratio can serve as a promising indicator of the disk's C/O ratio in the HD\,163296. Overall, the disk of HD\,163296 provides a favorable environment for the detection of sulfur-containing molecules.
}

   \authorrunning{Rong Ma }            
   \titlerunning{Investigating Sulfur Chemistry in the HD\,163296 disk}  

   \maketitle

%
%
\section{Introduction}           
\label{sect:intro}

A protoplanetary disk (PPD) is a crucial intermediate stage in the evolution from an interstellar molecular cloud to a planetary system (\citealt{2021ApJS..257...12L}). The PPDs serve both as conduits of the inherited interstellar and protostellar organic chemistry to planets and planetesimals and as active producers of new organic molecules using the disk inorganic and organic C reservoirs (\citealt{2023ARA&A..61..287O}). Along with the well-studied C, N, and O chemistry, S-containing molecules have also been observed in PPDs (e.g., \citealt{2013A&A...549A..92G},  \citealt{2015ApJ...799..204C}, \citealt{2018ApJ...857...69B}, \citealt{2019ApJ...876...25B}, \citealt{2021ApJS..257....2C}, ). Sulfur is one of the most abundant elements in the interin protoplanetary disks medium (\citealt{2009The}) and plays crucial roles in prebiotic chemistry (\citealt{2015Formation}) and planetary habitability (\citealt{2018AsBio..18.1023R}). However, sulfur chemistry is still poorly understood in the formation processes of low-mass stars and planets. Protoplanetary disks present an interesting experimental ground for investigating sulfur chemistry models. This is because each disk exhibits a wide range of unique environments. The study of S-containing species has proven invaluable in reconstructing the chemical history and kinetics of the studied objects (\citealt{2023NatAs...7..684K}). So far, the main S reservoirs have not yet been determined, and there is still much theoretical work to be done to determine the chemical pathways that produce the observed distribution of sulfur species (\citealt{2019ApJ...876...72L}).

So far, six sulfur-containing molecules have been detected in protoplanetary disks (\citealt{2021ApJS..257...12L}), including CS, H$_2$CS, H$_2$S, C$_2$S, SO, and SO$_2$ with CS the most easily detected one
 (\citealt{1997A&A...317L..55D}, \citealt{2011A&A...535A.104D}, \citealt{2018ApJ...864..133T}, \citealt{2019ApJ...876...72L}, \citealt{2021ApJS..257...12L}). Hydrogen sulfide (H$_2$S) has long been considered a significant sulfur reservoir but was recently found to be present in the disks (\citealt{2018A&A...616L...5P}). The H$_2$S/CS gas-phase column density ratio is typically around 1/20 (\citealt{2018A&A...616L...5P}). In the MWC\,480 disk, H$_2$CS were detected with a column density ratio $N$[H$_2$CS]/$N$[CS]$=$ 2/3, suggesting that a substantial part of the sulfur reservoir in disks is in organic form (\citealt{2021ApJS..257...12L}). The oxygen-sulfur compunds detected in the disks are SO and SO$_2$, only having been detected in very few sources (\citealt{2021A&A...651L...6B}, \citealt{2021ApJS..257...12L},  \citealt{2023A&A...669A..53B}). At present, some observations and theoretical studies of the disk indicate that the gas-phase C/O may vary with the radial position of the disk, and CS/SO is a probe to detect the carbon oxygen ratio of elements (\citealt{2021ApJS..257...12L}, \citealt{2023NatAs...7..684K}).

The HD\,163296 (MWC 275) system provides an ideal experimental platform to investigate chemical treatments within an original planetary disk. It is a solitary Herbig Ae pre-main sequence (PMS) star with a spectral type of A1 and an age of approximately 6 million years (\citealt{2015MNRAS.453..976F}). The star is encircled by a vast and luminous protoplanetary disk. The inclination of the disk is 46.7 degrees, with a position angle of 133 degrees (\citealt{2018ApJ...852..122H}). Only CS 2-1 has been observed in HD\,163596 (\citealt{2021ApJS..257...12L}). The observations by \cite{2021ApJS..257...12L} also encompassed the transition between C$_2$S and SO in \textcolor{black}{HD\,163296}, but only the upper limit was found.
 
In this paper, we present the detection of CS 7-6 toward HD\,163296, using the data obtained with Atacama Large Millimeter/submillimeter Array (ALMA) and illustrate our astrochemical modeling on HD\,163296 to study the formation and distribution of sulfur containing species. We discuss the CS 7-6 line detection toward HD\,163296 in Section \ref{sec:Observations}. In Section \ref{sec:Model}, we showcase grids of disk chemistry models tailored to the HD\,163296 disk. We also present the modeling results in Section \ref{sec:results}. Moreover, a detailed discussion is provided in Section \ref{sec:Discussion}, followed by a summary of our conclusions (Section \ref{sec:Conclusions}).

\section{Observations}
\label{sec:Observations}
\subsection{Observational Details} \label{subsec:Details}

The molecular line CS\,7\,-\,6 data was extracted from the ALMA archive (project code 2016.1.01086.S; PI: \textcolor{black}{Isella}). The total on-source integration time was 28 minutes. Forty-two antennas were used during the observations and the length of baselines ranges from 21.0\,m to 3.6\,km. For the calibrations, J1733\,-\,1304 was used as flux calibrator, J1751\,-\,1950 as phase calibrator and J1924\,-\,2914 as bandpass calibrator.

The data was calibrated by ALMA supplied pipeline scripts and then was imaged using Common Astronomy Software Application package (CASA) (\citealt{2007ASPC..376..127M}). 
The continuum  was subtracted towards the uv data. The line-only uv data was cleaned using “tclean” with “multiscale” algorithm ({\tt\string deconvolver=``multiscale''}) and a Briggs robustness parameter of 0.5, resulting in a spatial resolution of $\sim $0.066$^{''}$.
During the imaging, a Keplerian mask was used. The Keplerian mask was built based on the HD\,163296 disk and star parameters used in \cite{2021ApJS..257....2C}. For the spectral window including CS\,7\,-\,6, the bandwidth is
1.875\,GHz and the channel number is 1920, resulting in a channel width  976.562\,kHz, corresponding to a velocity width 0.85\,km\,s$^{-1}$ at 342.883\,GHz. The noise level of the line cube is 1.9\,mJy\,beam$^{-1}$ per channel. \textcolor{black}{The maximum recoverable scales of the observations is 0.953$^{''}$}

\subsection{Observational results} \label{subsec:Observational results}
\begin{figure}
   \centering
   \includegraphics[width=150mm]{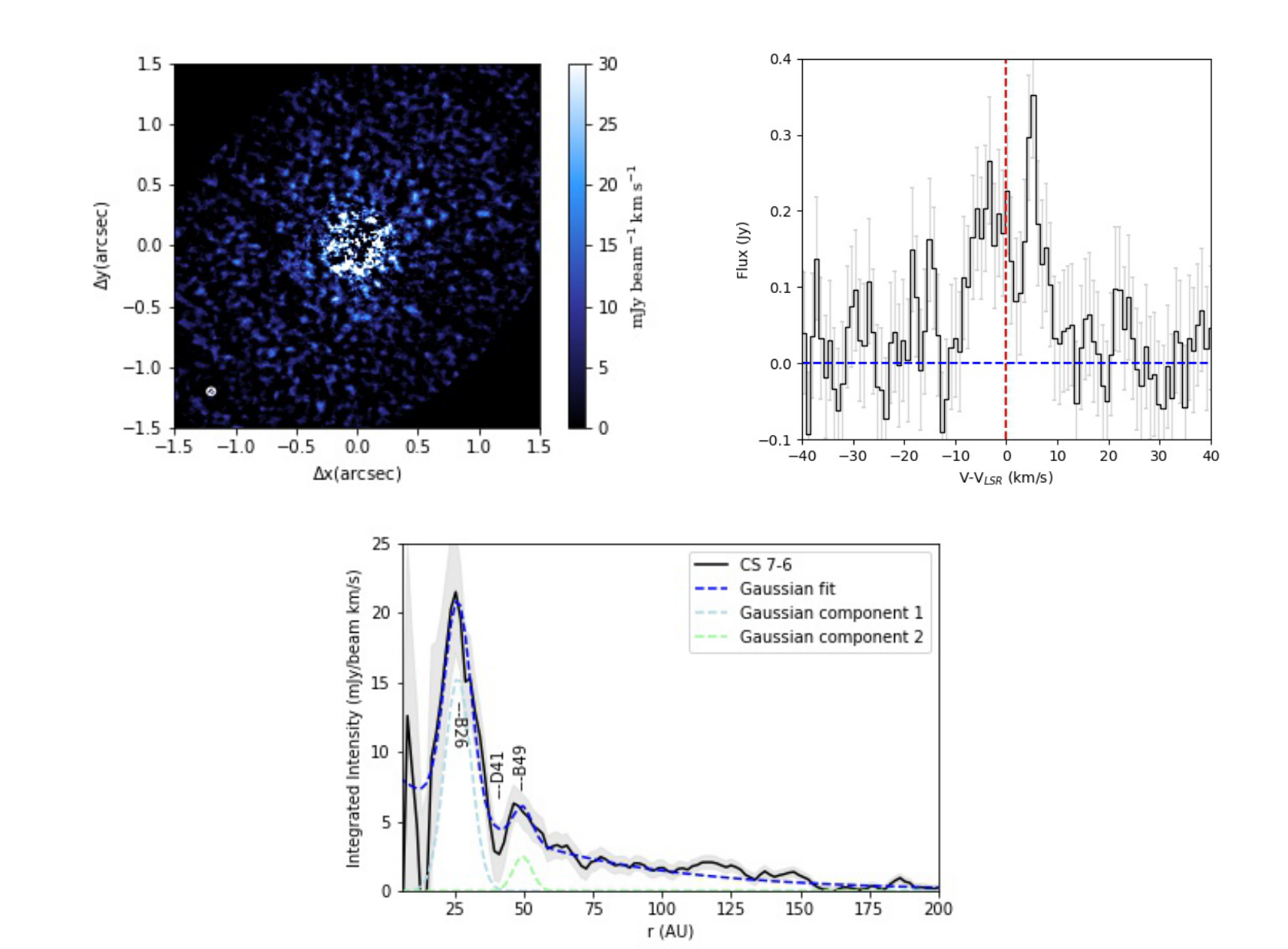}
   \caption{First row: Zeroth-moment maps and integrated intensity spectra of CS 7-6 for the HD\,163296 disk, from left to right. Second row: radial intensity profiles of CS 7-6. The black solid line represents the radial profile of the observed data. Blue dashed lines represent composite fitting, while individual Gaussian contours are displayed as light blue dashed lines. Chemical substructures are marked by solid and dotted arcs, indicating bright and dark features, respectively. The gray area represents 1\,$\sigma$ uncertainty.}
   \label{fig:1}
   \end{figure}

Figure \ref{fig:1} displays the integrated intensity (zeroth-moment) maps of the spatially resolved observations of the CS 7-6 rotational transition toward the HD\,163296. To build these maps, we used the Python package {\tt\string bettermoments} (\citealt{2018RNAAS...2..173T}) applied. We used a hybrid mask combining a Keplerian mask and 1$\sigma$ clip to mask any pixels below this threshold. Figure \ref{fig:1} also shows the radial profile of the shifted and stacked spectra using the {\tt\string radial profile} function in {\tt\string GoFish} (\citealt{2019JOSS....4.1632T}). Considering the disk physical parameters (i.e., disk inclination, disk position angle, mass of the central star, and distance) listed in Table \ref{tab:1}. 

\begin{figure}
   \centering
   \includegraphics[width=100mm]{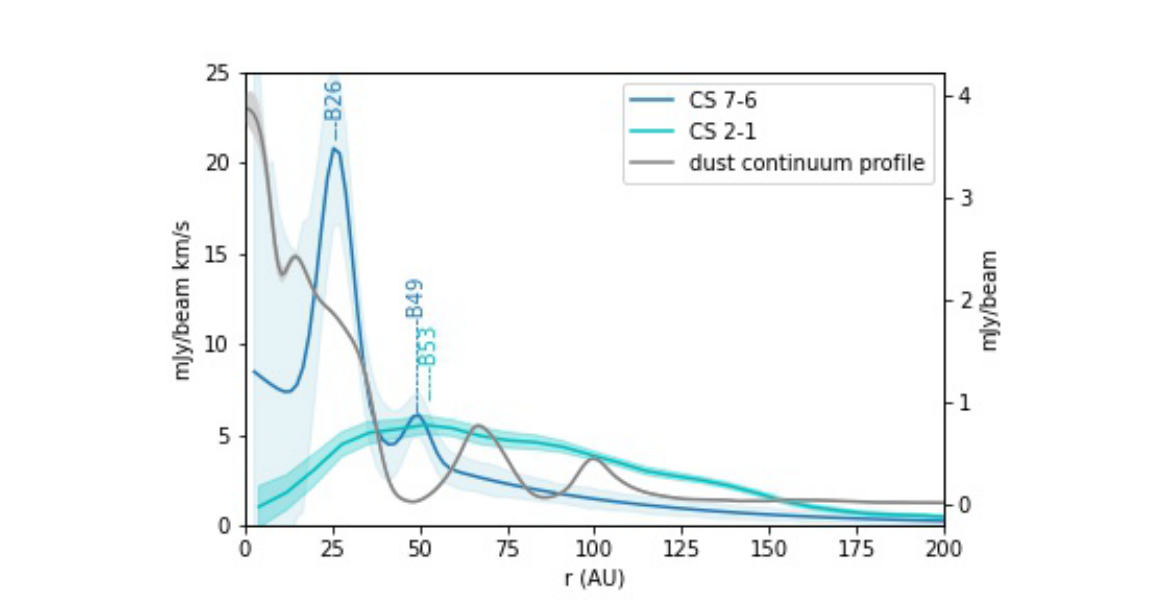}
   \caption{CS 2–1 and CS 7–6 radial intensity profiles. CS 2–1 line profile data from MAPS data (\citealt{2021ApJS..257....1O}), as described in \cite{2021ApJS..257....3L}. The solid grey line is the 1.25 mm dust continuum profile, data source from the ALMA DSHARP project (\citealt{2018ApJ...869L..41A}), as described in \cite{2018ApJ...869L..49I}.}
   \label{fig:2}
   \end{figure}

Based on the radial intensity profiles of CS 7-6 emission in HD\,163296, we discovered the presence of a central hole, similar to the CS 2-1 emission reported by \cite{2021ApJS..257...12L}. Following the method proposed by \cite{2018ApJ...869L..42H}, we modeled the radial profile of CS 7-6 as a sum of one or more Gaussian components using the Levenberg-Marquardt minimization implementation in {\tt\string LMFIT} (\citealt{2020zndo...3814709N}). We found that CS 7-6 emission has two bright rings with centers ranging from $\sim$26 to $\sim$49\,AU. In addition, a gap can be seen at $\sim$41\,AU. \textcolor{black}{By comparing the radial intensity distributions of the CS 2-1 emission and the CS 7-6 emission (see Figure \ref{fig:2}), it is observed that the location of one of the CS 7-6 ring (B49) is very close to the location of the CS 2-1 ring (B53, as described in \cite{2021ApJS..257....3L}). But the CS 7-6 emission has a protruding ring in the center hole range of the CS 2-1 emission. In addition, we can observe the different morphology of the CS 7-6 radial intensity profile compared to the 1.25\,mm dust continuum profile.}

The CS 7-6 spectra are also depicted in Figure \ref{fig:1} for HD\,163296 disk, showing a typical double-peaked profile indicative of the Keplerian rotation of the disk. The detection is made at a 24\,$\sigma$ confidence level, with a peak flux of 104\,mJy\,beam$^{-1}$\,km\,s$^{-1}$, and rms of 4.3\,mJy\,beam$^{-1}$ km\,s$^{-1}$ as measured from the emission-free regions of the integrated intensity map. The disk-integrated flux densities, $S_{\rm v} \Delta_{\rm v}$, derived from this work are listed in Table \ref{tab:Observations}.

\begin{table}
\bc
\begin{minipage}[]{100mm}
\caption[]{Disk characteristics of the HD\,163296.\label{tab:1}}\end{minipage}
\setlength{\tabcolsep}{2.5pt}
\small
 \begin{tabular}{ccccccccccc}
  \hline\noalign{\smallskip}
Source&Type& Dist. &Incl &PA  &$\rm L_\ast$ &Age  &$\rm M_\ast$ & $T_{\rm eff}$&vsys \\ 
& & (pc)&(deg)&(K)&($\rm L_\odot$) &(Myr)  &($\rm M_\odot$) &(K)&(km s$^{-1}$)\\
  \hline\noalign{\smallskip}
HD\,163296 &A1$^1$ &101$^3$&46.7$^2$&133.3$^2$&17$^1$&6$^1$&2$^5$&9332$^4$&5.8$^5$\\
  \noalign{\smallskip}\hline
\end{tabular}
\ec
\tablecomments{0.66\textwidth}{References. (1) \cite{2015MNRAS.453..976F}; (2) \cite{2018ApJ...852..122H}; (3) \cite{20181G}; (4) \cite{2018ApJ...869L..41A}; (5) \cite{2019Natur.574..378T}.}
\end{table}

\begin{table}
\bc
\begin{minipage}[]{100mm}
\caption[]{List of Observations (Molecular Data from CDMS).\label{tab:Observations}}\end{minipage}
\setlength{\tabcolsep}{2.5pt}
\small
 \begin{tabular}{ccccccccc}
  \hline\noalign{\smallskip}
Species&Transition&  Frequency &E$_u$ &log$_{10}($$A_{\rm ij}$)  &rms$_{\rm chan}$&$R_{\rm max}$  &$S_v \Delta _v$($R_{\rm max}$) \\ 
& & (GHz)&(K)&(s$^{-1}$)&(mJy beam$^{-1}$) &(arcsec)  &(Jy beam$^{-1}$) \\
  \hline\noalign{\smallskip}
CS&7-6&342.8828&65.8&-3.0773&1.9&2.000$\pm$0.017&4.95$\pm$0.32 \\
  \noalign{\smallskip}\hline
\end{tabular}
\ec
\tablecomments{0.76\textwidth}{\begin{itemize}\item{a. \href{https://cdms.astro.uni-koeln.de/cdms/portal}{https://cdms.astro.uni-koeln.de/cdms/portal}}. 
\item{b. $R_{\rm max}$ stands for the outer radius of the molecular line emission, where 95$\%$ of the cumulative flux from the radial profiles is contained. The uncertainty is 1$\sigma$ error.} 
\item{c. $S_v \Delta _v$($R_{\rm max}$) corresponds to the flux density integrated out to the outer radius $R_{\rm max}$ of the molecular line emission.}\end{itemize} }   
\end{table}

\section{Models} \label{sec:Model}
\subsection{Chemical models} \label{subsec:chemical}

In this study, we use the Nautilus three-phase model (\citealt{Wakelam+etal+2016}), which includes the gas-phase, dust surface (the top two layers of the ice mantles), and the inner ice mantles. The chemical network used in our study is derived from \cite{2017MNRAS.469..435V}, which improves the understanding and representation of the sulfur chemical network. \textcolor{black}{The chemical network contains 962 species and 10913 reactions, including 7040 gas-phase reactions and 3873 surface reactions.} The network includes various processes such as cosmic ray desorption, photodissociation, photoionization, and adsorption/desorption between gas-phase and ice mantle species. We also utilized a constant universal photodesorption yield for all molecules.

To understand the chemistry of the disk during planetary formation, we started from the molecular cloud stage and ran dense cloud chemistry models for 1 million years. The physical conditions we considered included a grain and gas temperature of 10\,K, the gas density of $\rm n_H= 2\times10^4\,cm^{-3}$, the visual extinction of 15\,mag and the cosmic-ray ionization rate of $\rm 1.3\times10^{-17}\,s^{-1}$. The initial abundance used can be found in Table \ref{tab:4}. The composition of the molecular cloud at the end of this stage was taken as the initial condition for the disk.

\subsection{Disk physical models}
In the second stage, we simulated the chemical composition of the disk for $10^7$ years using a cosmic-ray ionization rate of $\rm 1.3\times10^{-17}\,s^{-1}$. 

We assume that the disk model is axisymmetric and includes gas, small dust, and large dust components. There is spatial coupling between the gas and small dust populations. The mass surface distribution of gas and small dust follows a self-similarly viscous disk model, as described by \cite{1974MNRAS.168..603L} and \cite{2011ApJ...732...42A}. The disk structure is fully parameterised, with a surface density that follows the standard form of a power law with an exponential taper:
\begin{equation}
  \rm{\Sigma(r)={\Sigma_c}({\frac{r}{R_c}})^{-\gamma}\exp[{-(\frac{r}{R_c}})^{2-\gamma}]},
\end{equation}
where $\rm R_c$ is the characteristic scaling radius, $\rm \Sigma_c$ is surface density at characteristic radius, and $\gamma$ is the gas surface density exponent.
The vertical density structure is assumed to be a gaussian function characterized by a scale height H(R) that is a powerlaw function of radius:
\begin{equation}
  \rho(r,z)=f_{\rm i}\frac{\Sigma(r)}{\sqrt{2\pi}H_{\rm i}(r)}\exp[{-\frac{1}{2}(\frac{z^2}{H_{\rm i}(r)})}].
\end{equation}
The scale height is then given by
\begin{equation}
  H_{\rm i}(r)=\chi H_{100}{\frac{r}{100/AU}}^\psi,
\end{equation} 
where $f_{\rm i}$ is the mass fraction of each mass component, $H_{100}$ is the scale height at 100\,AU, and $ \psi $ is a parameter that characterizes the radial dependence of the scale height.  Here we fix $\chi=1$ for the gas and the small grain population and \textcolor{black}{$\chi=0.2$} for the large grain population (\citealt{2011ApJ...732...42A}). Both dust populations follow a Mathis Rumpl Nordsieck grain distribution $ n(a)\propto a^{-3.5}$ (\citealt{1977ApJ...217..425M}). The following description of the dust grain population used in this study is adopted from \cite{2021ApJS..257....5Z}. For a given dust density structure , we calculate the dust temperature structure using the Monte Carlo radiative transfer code RADMC3D (\citealt{2012ascl.soft02015D}).

For gas temperature, we used a two-layer model similar to the one proposed by \cite{2003A&A...399..773D}, but later modified by \cite{2021ApJS..257....4L} with different connection terms. Assuming that the midplane temperature $T_{\rm mid}$ and atmospheric temperature $T_{\rm atm}$ have power law distributions with slopes $q_{\rm mid}$ and $q_{\rm atm}$, respectively:
\begin{equation}
 T_{\rm atm}=T_{\rm atm,R_0}(\frac{r}{R_0})^{-q},
\end{equation}
\begin{equation}
 T_{\rm mid}=T_{\rm mid,R_0}(\frac{r}{R_0})^{-q}.
\end{equation}
Between the midplane and atmosphere, the temperature is smoothly connected using a tangent hyperbolic function
\begin{equation}
 T^{4}(r,z)=T^4_{\rm mid}(r)+1/2[1+tanh(\frac{z-\alpha z_{\rm q}(r)}{z_{\rm q}(r)})]T^4_{\rm atm}(r),
\end{equation} 
where $z_{\rm q}(r)=z_0(r/100 AU)^\beta$. We note that the parameter $\alpha$ determines the height at which the transition in the tanh vertical temperature profile takes place, while $\beta$ represents the variation of transition height across the radius. There are a total of the following parameters $T_{\rm mid}$, $T_{\rm atm}$, $q_{\rm mid}$, $q_{\rm atm}$, $\alpha$, $\beta$, $z_0$ whose values are from the \cite{2021ApJS..257....4L}. 

The ultraviolet (UV) flux at a specific radius is determined by the sum of photons emitted directly from the star and those that are scattered downward by small dust particles in the disk's atmosphere (\citealt{Wakelam+etal+2016}). This relationship is mathematically represented by an equation,
\begin{equation}
  f_{\rm uv}=\frac{0.5 f_{\rm uv,R_0}}{(\frac{r}{R_0})^2+(\frac{4H}{R_0})^2}.
\label{eq:UV}
\end{equation}

\begin{figure}
   \graphicspath{{figure/}}
   \centering
   \includegraphics[width=15cm]{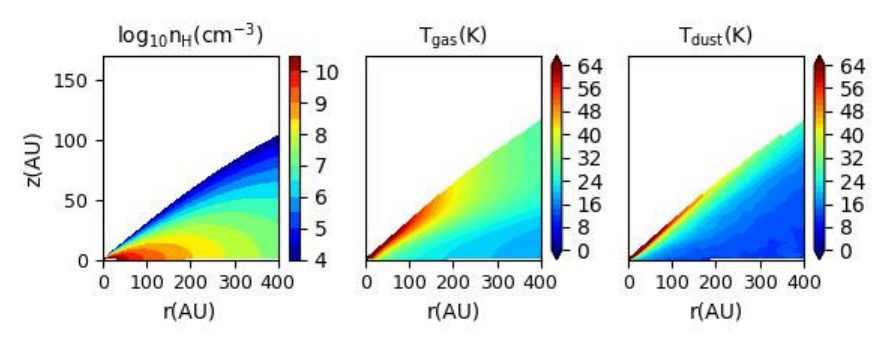}
   \caption{Calculated number density of H, gas temperature (middle) and dust temperature as a function of radius $r$ and height $z$ for the HD\,163296 disk.  }
   \label{fig:3}
   \end{figure}

The physical disk parameters used in our model are given in Table \ref{tab:1} and Table \ref{tab:models}. Disk density and temperature profiles around HD\,163296 are shown in Figure \ref{fig:3}. 

\begin{table}
\bc
\begin{minipage}[]{100mm}
\caption[]{Physical parameters of models.\label{tab:models}}\end{minipage}
\setlength{\tabcolsep}{2.5pt}
\small
 \begin{tabular}{ccc}
  \hline\noalign{\smallskip}
Parameters &  
&References\\
  \hline\noalign{\smallskip}
Surface density of gas at $R_{\rm c}$ ($\rm g\,cm^{-2}$)& 8.8 & \citealt{2021ApJS..257....5Z}\\
Surface density of the small-grain population at $R_{\rm c}$ ($\rm g\,cm^{-2}$) &  1.3e-02 & \citealt{2021ApJS..257....5Z}\\
Characteristic radius of the gas and the small-grain population $R_{\rm c}^g$ (AU) &  165 & \citealt{2021ApJS..257....5Z}\\
Surface density exponent $\gamma$ &0.8 & \citealt{2021ApJS..257....5Z}\\
Scale height at 100 AU & 8.4 & \citealt{2021ApJS..257....5Z}\\
Power index of the radial dependence of scale height $ \psi $  & 1.08 & \citealt{2021ApJS..257....5Z}\\
The inner edge of the disk model $r_{\rm in}$ (AU)& 0.45 & \citealt{2021ApJS..257....5Z}\\
The outer edge of the disk model $r_{\rm out}$ (AU)& 600 & \citealt{2021ApJS..257....5Z}\\
Midplane temperature at   $R_{\rm c}$   (K) & 63 &\citealt{2021ApJS..257....4L}\\
Atmosphere temperature at  $R_{\rm c}$   (K) &27 &\citealt{2021ApJS..257....4L} \\
Midplane Temperature power-law index: $q_{\rm atm}$ & 0.5  &\citealt{2021ApJS..257....4L}\\
Atmosphere Temperature power-law index: $q_{\rm mid}$ & 0.5  &\citealt{2021ApJS..257....4L}\\
$\alpha$ & 3.01 &\citealt{2021ApJS..257....4L}\\
$\beta$ &0.42 &\citealt{2021ApJS..257....4L}\\
$z_0$ (AU)&9 &\citealt{2021ApJS..257....4L}\\
$f_{\rm i}$ (for large grain population) & 0.80 & \citealt{2021ApJS..257....5Z} \\
$f_{\rm uv,R_0}$(in Draine (1978)’s units) & 8500 & This work \\

  \noalign{\smallskip}\hline
\end{tabular}
\ec
\end{table}

\begin{table}
\bc
\begin{minipage}[]{100mm}
\caption[]{Initial elemental abundances.\label{tab:4}}\end{minipage}
\setlength{\tabcolsep}{2.5pt}
\small
 \begin{tabular}{ccc}
  \hline\noalign{\smallskip}
Species &$n_i/n_H $ & Reference\\
  \hline\noalign{\smallskip}
$\rm H_2$                & 0.5                     &    \\
$\rm He$                 & $\rm 9.0\times10^{-2}$      & \citealt{Guilloteau+etal+2011}    \\
$\rm C^+$                & $\rm 1.7\times10^{-4}$      & \citealt{Pitu+etal+2007}    \\
$\rm N$                  & $\rm 6.2\times10^{-5}$      & \citealt{Pitu+etal+2007}  \\
$\rm O $                 & $\rm 2.4\times10^{-4}$      & \citealt{201161H}    \\
$\rm S^+$                & $\rm 8.0\times10^{-9}$      & \citealt{Pizzo+etal+2011}    \\
$\rm Si^+$               & $\rm 8.0\times10^{-8}$      & \citealt{Pizzo+etal+2011}   \\
$\rm Fe^+$               & $\rm 3.0\times10^{-9}$      & \citealt{Pizzo+etal+2011}    \\
$\rm Na^+$               & $\rm 2.0\times10^{-9}$      & \citealt{Pizzo+etal+2011}     \\
$\rm Me^+$               & $\rm 7.0\times10^{-9}$      & \citealt{Pizzo+etal+2011}    \\
$\rm P^+$                & $\rm 2.0\times10^{-10}$     & \citealt{Pizzo+etal+2011}   \\
$\rm Cl^+$               & $\rm 1.0\times10^{-9}$      & \citealt{Pizzo+etal+2011}    \\
$\rm F$                  & $\rm 6.7\times10^{-9}$      & \citealt{2015AA}     \\
  \noalign{\smallskip}\hline
\end{tabular}
\ec
\end{table}

\section{Modeling results} \label{sec:results}

\begin{figure}
   \graphicspath{{figure/}}
   \centering
   \includegraphics[width=150mm]{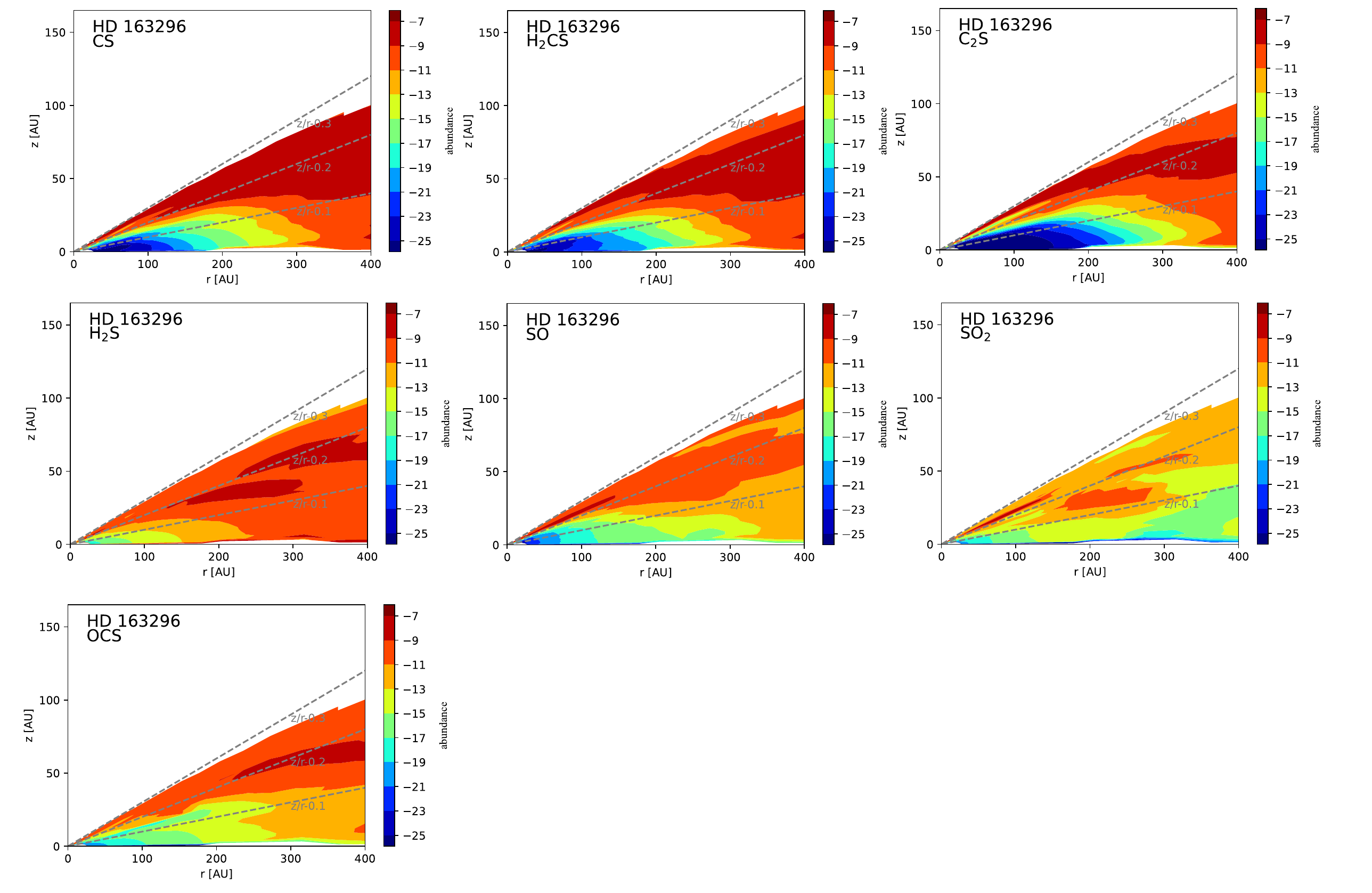}
   \caption{Calculated abundance of sulphur molecules in the gas-phase as a function of radius $r$ and height over radius $z$ for the HD\,163296 disk.}
   \label{fig:4}
   \end{figure}

\begin{figure}
   \graphicspath{{figure/}}
   \centering
   \includegraphics[width=150mm]{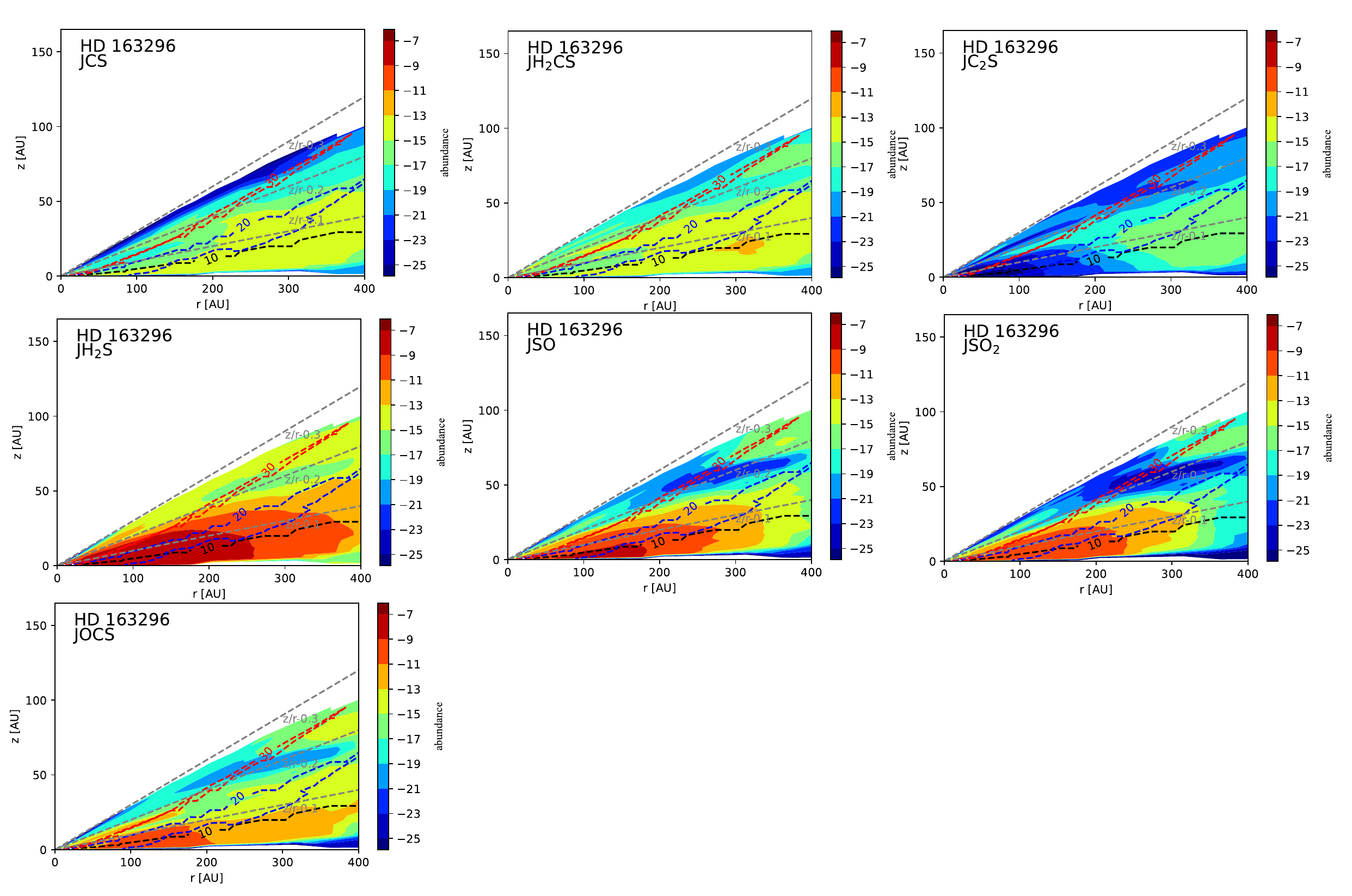}
   \caption{Calculated abundance of sulphur molecules on the ice surface  as a function of radius $r$ and height over radius $z$ for the HD\,163296 disk. The black, blue and red dotted lines in the figure are the dust temperatures for the 10 K, 20 K and 30 K contours, respectively.}
   \label{fig:5}
   \end{figure}

In this section, we present the calculated abundance distribution of S-bearing molecules in the HD\,163296 disk model in Figure \ref{fig:4} and \ref{fig:5}. Then we discussed the sources and formation pathways of these molecules. \textcolor{black}{In the following, J is added before the molecular formula to represent the molecules on the ice surface.}

\textbf{CS and JCS}: From Figure \ref{fig:4}, it can be found that CS mainly exists in the gas-phase between z/r$ \sim $0.1 and z/r$ \sim $0.3. In this region, the gas-phase abundance of CS can reach 10$^{-7}$. In the area near the midplane (z/r \textless0.1), the abundance of CS is relatively low. As the distance from the central star increases, the abundance of CS increases from 10$^{-25}$ to 10$^{-9}$ in the region near the midplane. CS is produced through the bimolecular reaction of H + HCS$\rightarrow$CS + H$_2$, and the electron recombination of HCS$^+$ and H$_3$CS$^+$ in the region of the disk where z/r\textgreater0.1 and z/r\textless0.3. Among them, HCS$^+$ and H$_3$CS$^+$ are generated through rapid ion neutral reactions between S$^+$ and small hydrocarbon CH$_4$. Subsequently, H$_3$CS$^+$ recombines with electrons to form neutral substances containing S, such as CS and HCS. In the midplane of the disk, CS is freeze-out onto grain surfaces. In addition, the main destruction pathway of CS is through reactions with H$_3^+$, HCO$^+$ and OH. JCS storage is concentrated in an area near the midplane of the disk, with an abundance of 10$^{-15}$-10$^{-13}$. 

\textbf{H$_2$CS and JH$_2$CS}: The abundance distribution of H$_2$CS trend is similar to CS and is mainly found at intermediate heights (0.1\textless z/r\textless0.25) in the HD\,163296 disk. The abundance of H$_2$CS can reach 10$^{-25}$-10$^{-9}$ at intermediate heights. H$_2$CS mainly consists of gas-phase reactions between S atoms and CH$_3$, as well as dissociation and recombination of electrons H$_3$CS$^+$. Near the midplane, H$_2$CS is formed by chemical desorption in the gas-phase through surface reactions JS + JCH$_3\rightarrow$H$_2$CS + H and JH + JHCS$\rightarrow$H$_2$CS. H$_2$CS mainly reacts with protonated ions to destroy (H$^+$, H$_3^+$, HCO$^+$) and freezes on the surface of grains near the midplane. JH$_2$CS abundance is 10$^{-15}$-10$^{-13}$ in most regions with z/r\textless0.2, similar to JCS.

\textbf{C$_2$S and JC$_2$S}: The distribution of C$_2$S is similar to that of CS and H$_2$CS, but the gas-phase abundance can only reach 10$^{-25}$-10$^{-9}$ at the narrow middle height (0.15\textless z/r\textless0.25). C$_2$S is formed partly through gas-phase neutral reactions between S atoms and C$_2$H, and partly through the electronic
dissociative recombination of H$_3$CS$^+$ and H$_2$CS$^+$. The destruction reaction of C$_2$S is identical to that of H$_2$CS, which explains the similarity in their distribution.
JC$_2$S exhibits lower abundance compared to JH$_2$CS and JCS due to its involvement in hydrogenation reactions, resulting in the production of longer S-containing compounds (JHCCS).

\textbf{H$_2$S and JH$_2$S}: The abundance of H$_2$S in a localized area with a radial distance of r\textgreater150\,AU and a height of z/r-0.2 is 10$^{-7}$. H$_2$S part of in the gas-phase is achieved through recombination of electrons H$_3$CS$^+$ is generated, while another part is generated on dust particles and released into the gas-phase through Photodesorption by UV photons and cosmic-ray evaporation. We found that JH$_2$S mainly exists in ice near the midplane (z/r\textless1), and the abundance of H$_2$S can reach 10$^{-9}$-10$^{-7}$. H$_2$S freezes on the surface of grains in this area. JH$_2$S is produced through the reaction JH + JHS$\rightarrow$JH$_2$S.

\textbf{SO and JSO}: The abundance of SO is 10$^{-11}$ to 10$^{-9}$ between z/r$ \sim $0.1 and z/r$ \sim $0.3. In the midplane, for r below 100\,AU, the abundance of SO is less than 10$^{-17}$. However, as the radius increases, the abundance progressively rises, eventually reaching 10$^{-11}$. SO is generated in the gas-phase through the reaction of S and OH at 0.1\textless z/r\textless0.3. The destruction mechanism for SO resembles that of H$_2$CS. SO are destroyed by H$_3^+$ and HCO$^+$. In the region near the midplane, SO freeze onto dust particles. This effect is especially prominent for r below 200\,AU. In this zone, the abundance of JSO can reach 10$^{-7}$.

\textbf{SO$_2$ and JSO$_2$}: In localized regions (150\,AU\textless r\textless250\,AU) with z/r ranging from 0.1 to 0.2, the abundance of SO$_2$ is 10$^{-11}$-10$^{-9}$. Its formation in the HD\,163296 disk occurs through the reaction: SO + OH$\rightarrow$SO$_2$ + H. JSO$_2$ exhibits a distribution similar to JSO since it is formed via surface reactions between JSO and JO.

\textbf{OCS and JOCS}: The abundance of OCS can reach 10$^{-7}$ near r\textgreater200\,AU and z/r$ \sim $0.2. OCS is frozen onto the ice in regions close to the midplane. We have found that OCS is formed through reactions with S and CO, as well as reactions with CS and OH in the gas-phase. The abundance of JOCS can reach 10$^{-9}$ near the midplane. In regions, JOCS can be formed through reactions with JS and JCO.

\section{Discussion} \label{sec:Discussion}
\subsection{Comparison of model-simulated column densities and observationally deduced fractional column densities}
This section presents the column densities of the species investigated using our model, as depicted in
Figure \ref{fig:6}. We compared the model results against existing \textcolor{black}{observationally deduced fractional column densities}, summarized the similarities and discrepancies.
\textcolor{black}{\cite{2021ApJS..257...12L} derived the disk-integrated column density of the CS in the HD\,163296 disk by observed the CS 2-1. This estimation was made assuming optically thin lines and local thermal equilibrium (LTE), with excitation temperatures in the range of 10-30\,K. \cite{2021ApJS..257...12L} also covered transitions of C$_2$S and SO in their study, and they estimated the upper limits of the SO and C$_2$S column densities.}
These results are depicted in the shaded region of Figure \ref{fig:6}. In this section, our main focus is on simulating the column densities at a carbon-to-oxygen ratio of 0.7. 

In Figure \ref{fig:6}, it is observed that the vertical column density of CS in our simulated HD\,163296 increases as it moves away from the star. When r is greater than 100\,AU, the column density of CS gradually becomes flat, reaching 10$^{13}$\,cm$^{-2}$. Our model results are consistent with the \textcolor{black}{observationally deduced fractional CS column density} (Within an order of magnitude range). \textcolor{black}{In addition, we found a protruding ring in the CS 7-6 radial profile (B26), while the chemical modeling results show a protrusion in the radial column density of CS at 26-32 AU (using different C/O ratio models).} However, it should be noted that the column density obtained by \cite{2021ApJS..257...12L} does not represent a radial variation. To achieve a more accurate comparison, it is necessary to observe more lines of CS. The vertical column density of C$_2$S increases with radius. It reaches 10$^{12}$\,cm$^{-2}$ in the outer disk region. \textcolor{black}{Comparison with the observationally deduced upper limits on C$_2$S column densities in \cite{2021ApJS..257...12L} shows that our simulated C$_2$S vertical column density is lower}. For the vertical column density of SO, we simulated its decrease from 10$^{14}$\,cm$^{-2}$ to 10$^{12}$\,cm$^{-2}$ as moved away from the central star. The simulated vertical column density of SO falls below the \textcolor{black}{observationally deduced upper limits on column densities} when r exceeds 180\,AU.

\subsection{Impact of the C/O Ratio}
\begin{figure}
   \graphicspath{{figure/}}
   \centering
   \includegraphics[width=150mm]{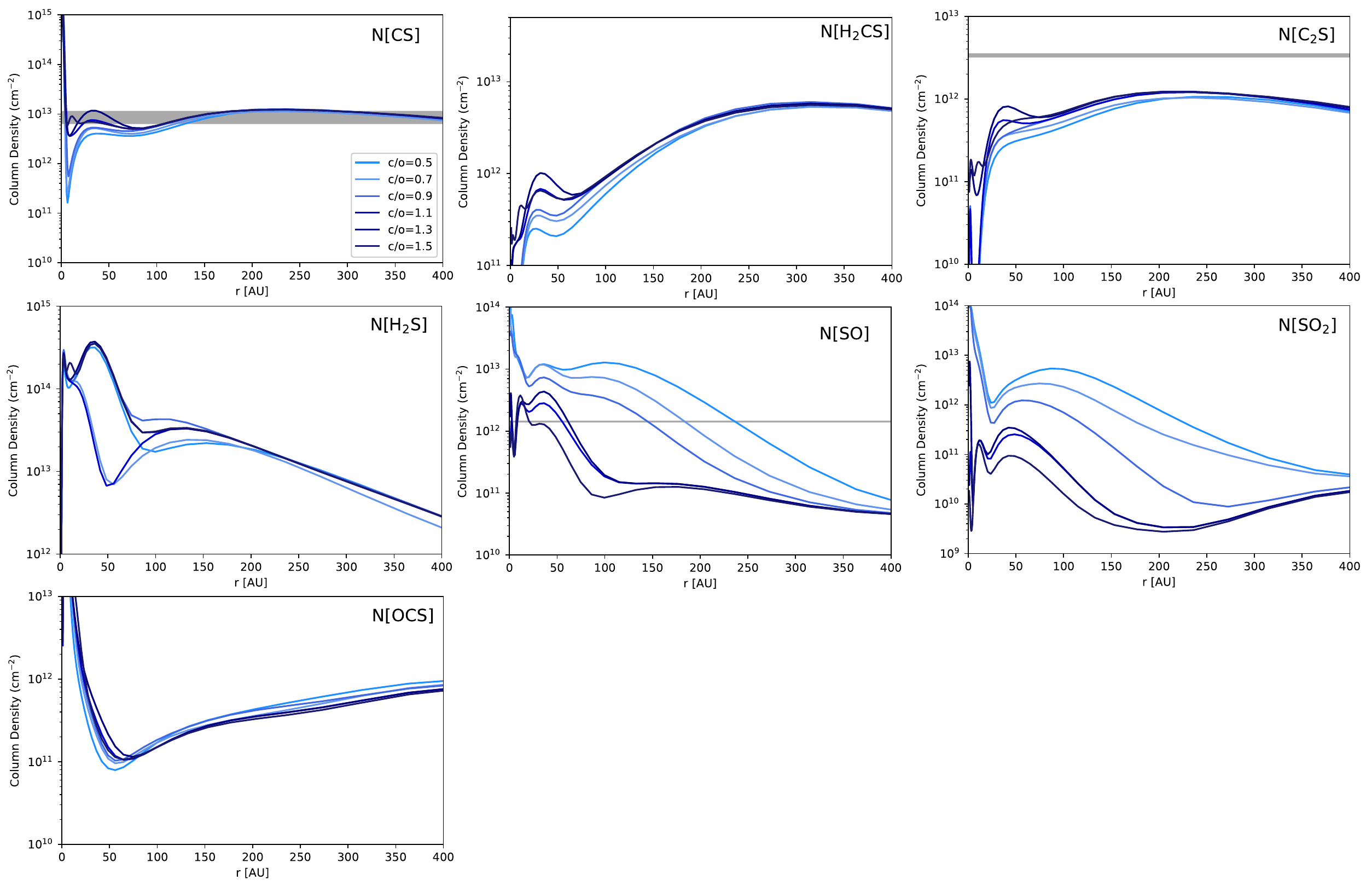}
   \caption{The column densities of the. CS, H$_2$CS, C$_2$S, H$_2$S, SO, SO$_2$, and OCS models were tuned to the HD\,163296 disk, vertically integrated from the upper layer of the disk to the midplane. The column density of the model is influenced by the investigation of C/O (C/O ratio ranging from 0.5 to 1.5). The gray shaded area represents the column density and upper limit calculated from the observed values, pointing towards the HD\,163296 disk (\citealt{2021ApJS..257...12L}).}
   \label{fig:6}
   \end{figure}

With the improvement of telescope resolution, there is an increasing trend on the observation and theoretical study of molecules within the protoplanetary disks. Specifically, there is a growing interest in understanding how the gaseous C/O ratio varies with the radial position and between different disks. In order to simulate the impact of C/O ratio on the column density of sulfur-containing molecules, we varied the C/O ratio from 0.5 to 1.5 (C abundance fixed). Figure \ref{fig:6} illustrates the influence of the gas-phase C/O ratio on the column densities of CS, H$_2$CS, SO, SO$_2$, C$_2$S, and H$_2$S in the HD\,163296 disk.

For the CS, H$_2$CS, and C$_2$S, we found that a higher C/O ratio leads to higher column densities for them. Because CS, H$_2$CS, and C$_2$S are driven by the reaction of S with small hydrocarbon (CH$_2$ and CH$_3$), this results in the formation of carbonated S-ions (HC$_2$S$^+$ and HC$_3$S$^+$), which subsequently undergo electron recombination to form neutral species containing sulfur (more details in Section \ref{sec:results}). As the C/O ratio increases, the abundance of hydrocarbons also increases, leading to higher column densities of CS, H$_2$CS, and C$_2$S.

For SO and SO$_2$, their column densities are highly sensitive to the C/O ratio. When the C/O ratio changes from 0.5 to 1.5, the largest change in column density for SO and SO$_2$ is a decrease of four orders of magnitude. This is because when there is a lack of oxygen, the abundance of the precursor OH for SO decreases, resulting in low column densities of SO and SO$_2$. In addition, we found that when the C/O ratio is less than 1.1, the overall trend of column density for SO and SO$_2$ decreases with increasing radius. When the C/O ratio is greater than 1.1, the column density of SO and SO$_2$ initially increases and then decreases with increasing radius. Due to observational limitations, we are unable to determine the best fitting model for the C/O ratio in the HD\,163296 disk. However, when C/O\textgreater1.1, the column densities of SO in more regions are lower than the observational upper limit of SO. Therefore, we speculate that the C/O ratio in the HD\,163296 disk may be greater than 1.1.

The column density of H$_2$S is sensitive to the C/O ratio in the region close to the central star (r\textless100\,AU). When the C/O ratio is 1.5, the column density of H$_2$S is highest. This is because a high C/O ratio increases the abundance of H$_3$CS$^+$, which subsequently reacts with electrons to generate H$_2$S in the gas-phase.

We found that when r is less than 50\,AU, the column density of OCS increases with a higher C/O ratio. This is because in this region, CO exists in the gas-phase, and OCS is primarily formed through reactions with S and CO. A higher C/O ratio leads to a higher abundance of CO, resulting in a larger column density of OCS. However, when r is less than 70\,AU, a higher C/O ratio leads to a lower column density of OCS. In this case, CO is frozen onto the icy surface, and the generation of OCS depends on the abundance of OH. A higher C/O ratio leads to a lower abundance of OH, resulting in a lower column density of OCS.

   \begin{figure}
   \graphicspath{{figure/}}
   \centering
   \includegraphics[width=150mm]{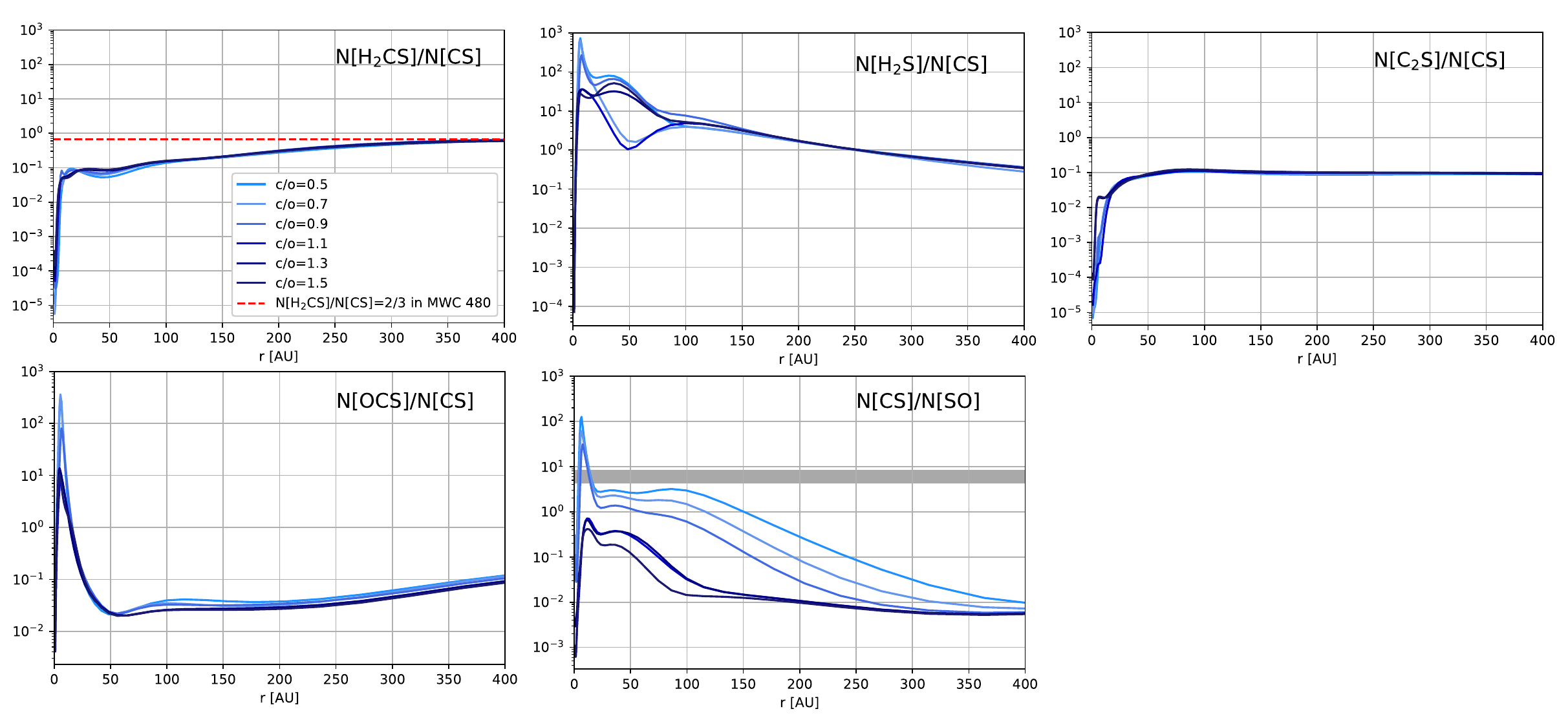}
   \caption{Calculated $N$[H$_2$CS]/$N$[CS], $N$[H$_2$S]/$N$[CS], $N$[C$_2$S]/$N$[CS], $N$[OCS]/$N$[CS], and $N$[CS]/$N$[SO] column density ratios for a grid of models tuned to the HD\,163296 disk investigating the impact of C/O ratios. The observed values of $N$[H$_2$CS]/$N$[CS] towards the MWC\,480 disk are from \cite{2021ApJS..257...12L}, represented by a red dashed line. $N$[CS]/$N$[SO] observations upper limit toward the HD\,163296 disk are indicated by the gray horizontal ower limits.}
   \label{fig:7}
   \end{figure}
   
\subsection{C/O ratio probe in the HD\,163296}

As shown in Figure \ref{fig:7}, our modeling results indicate that the column density of $N$[CS]/$N$[SO] changes by more than one order of magnitude when the C/O ratio varies from 0.5 to 1.5. This suggests that the ratio of $N$[CS]/$N$[SO] column densities is a promising probe for the disk C/O ratio in HD\,163296. \textcolor{black}{\cite{2021ApJS..257....7B} constrained the C/O ratio in HD 163296 by interpreting high-resolution C$_2$H observations. They found that a C/O ratio greater than 2 was required to match the high column density (\textgreater10$^{13}$ cm$^{-2}$) of C$_2$H in HD 163296 disks, but it is not obvious what causes the radial structures in the C$_2$H  column density.} \textcolor{black}{As the exoplanet community endeavors to measure the C/O ratio in planetary atmospheres, observational and theoretical studies of disks are increasingly focused on understanding how gas-phase C/O varies with radial position and between disks (\citealt{2023NatAs...7..684K}). Our modelling results show that $N$[CS]/$N$[SO] varies significantly as the C/O ratio varies in HD\,163296}. This may indicate that $N$[CS]/$N$[SO] is an ideal probe of the variation of gas-phase C/O with radial position in HD\,163296 disk. On the other hand, the column density ratios of $N$[H$_2$CS]/$N$[CS], $N$[C$_2$S]/$N$[CS], and $N$[OCS]/$N$[CS] remain largely unchanged with variations in C/O ratio, even in the presence of O-elements in OCS. Of particular interest is the $N$[H$_2$S]/$N$[CS] column density ratio, which, as observed from Figure \ref{fig:7}, exhibits significant changes with the C/O ratio when r is less than 100\,AU. This may imply that $N$[H$_2$S]/$N$[CS] can potentially trace the C/O ratio in the inner disk. However, further validation of these model results is required through more observations of diverse samples.

\subsection{Investigating the columns density ratio of sulfur-containing molecular in the HD\,163296}

Our simulation results show that $N$[H$_2$CS]/$N$[CS] within the range of 0.1 to 2/3 in HD\,163296. Previous observations of H$_2$CS in protoplanetary disks have been limited, with only a few disks detected thus far (\citealt{2021ApJS..257...12L}, \citealt{2020A&A...644A.120C}). \cite{2021ApJS..257...12L} detected H$_2$CS in the MWC\,480 disk, and their found indicate column density ratio of approximately 2/3 for $N$[H$_2$CS]/$N$[CS], suggesting a significant portion of the sulfur reservoir in the disk exists in organic form (i.e., C$_x$H$_y$S$_z$). Our calculations show that $N$[H$_2$CS]/$N$[CS] reaches a value of 2/3 for r larger than 300\,AU, similar to the observations in the MWC\,480 disk. This suggests the possibility of detecting H$_2$CS in the HD\,163296 disk. H$_2$S has been observed in several PPDs (\citealt{2021A&A...652A..46R}, \citealt{2022A&A...665A..61R}). \textcolor{black}{\cite{2021A&A...652A..46R} detected H$_2$S in several young stellar objects that are located in Taurus and indicated that the displayed value range for the $N$[H$_2$S]/$N$[CS] ratio was 0.12 to 0.38}. Our model shows that the $N$[H$_2$S]/$N$[CS] ratio is greater than 0.1 in most regions. This suggests that the disk of HD\,163296 is a favorable environment for detecting H$_2$S. The $N$[C$_2$S]/$N$[CS] ratio is mainly around 0.1 in most regions, similar to the ratio in the molecular clouds (\citealt{2018MNRAS.478.5514V}). The $N$[OCS]/$N$[CS] ratio is generally below 0.1 after r exceeds 20\,AU, indicating that OCS is difficult to detect in the HD\,163296 disk.

\section{Conclusions} \label{sec:Conclusions}
\textcolor{black}{We have analyzed the ALMA observations of CS 7-6 toward the HD\,163296 disk} and conducted astronomical chemical modeling on HD\,163296 to explore its sulfur chemistry. The main results of our study can be summarized as follows:
\begin{enumerate}
\item The CS 7–6 rotational transition was detected toward the HD\,163296 disk.
\item We conducted astrochemical modeling on HD\,163296 and compared the two-dimensional disk astrochemical model with \textcolor{black}{observationally deduced fractional column densities of CS, C$_2$S, and SO} in the HD\,163296 disk. We found that the CS and C$_2$S chemistry appears to be well understood, with the density of the CS simulation column showing good agreement with the observed column density calculated by LTE. The density of the C$_2$S simulation column falls below the upper limit of the observed values.
\item We investigated the influence of the C/O ratio on sulfur-containing molecules. We found that the column densities of SO and SO$_2$ are highly sensitive to the C/O ratio, and the column density of H$_2$S is particularly sensitive to the C/O ratio in the vicinity of the central star (r\textless100\,AU).
\item Using the astrochemical disk model, we discovered that the $N$[CS]/$N$[SO] ratio serves as a promising indicator of the C/O ratio in the HD\,163296 disk. Furthermore, the ratios of $N$[H$_2$CS]/$N$[CS], $N$[C$_2$S]/$N$[CS], and $N$[OCS]/$N$[CS] exhibit minimal variations with changes in the C/O ratio. Notably, $N$[H$_2$S]/$N$[CS] shows potential in tracking the C/O ratio of inner disks.
\item Our modeling results suggest that the disk of HD\,163296 is a promising environment for detecting sulfur-bearing molecules such as H$_2$CS and H$_2$S. 

\end{enumerate}

\begin{acknowledgements}
This work was funded by the National Natural Science Foundation of China (NSFC) under NSFC No.12373026, 11973075, 12203091, \textcolor{black}{12173075}. The National Key R$\&$D Program of China under grant No. 2022YFA1603103. The Natural Science Foundation of Xinjiang Uygur Autonomous Region of China (2022D01A156), the "Tianchi Doctoral Program 2021". Dalei li is supported by the Youth Innovation Promotion Association CAS.
This paper makes use of the following ALMA data: ADS/JAO.ALMA 2016.1.01086.S. ALMA is a partnership of ESO (representing its member states), NSF (the United States), and NINS (Japan), together with NRC (Canada), MOST and ASIAA (Taiwan), and KASI (the Republic of Korea), in cooperation with the Republic of Chile. The Joint ALMA Observatory is operated by ESO, AUI/NRAO, and NAOJ.
\end{acknowledgements}


\label{lastpage}
\end{document}